\documentclass[aps,preprint,showpacs,superscriptaddress]{revtex4-1}  
\usepackage{amssymb}
\usepackage{amsthm}
\usepackage{graphicx}
\usepackage{graphics}
\usepackage{bm}   
\usepackage{color}
\usepackage{amsmath}
\usepackage{balance}
\usepackage{subfigure}
\usepackage{relsize}
\usepackage{float}
\newcommand{\mathsym}[1]{{}}
\newcommand{\unicode}[1]{{}}
\usepackage{setspace}

\begin{document}
	
	\title{Quantum Squeezing Effects in Coupled van der Pol Oscillators}
	\author{M. Preethi}
	\affiliation{Department of Nonlinear Dynamics, Bharathidasan University, Tiruchirappalli - 620 024, Tamilnadu, India.}
	\author{M. Senthilvelan}
	\email[Correspondence to: ]{velan@cnld.bdu.ac.in}
	\affiliation{Department of Nonlinear Dynamics, Bharathidasan University, Tiruchirappalli - 620 024, Tamilnadu, India.}

\begin{abstract}
	Achieving synchronized quantum states within the quantum realm is a significant goal. This regime is characterized by restricted excitation occurrences and a highly nonclassical stable state of the self-oscillating system. However, many existing approaches to observe synchronization in this quantum realm face a major challenge: the influence of noise tends to overshadow the synchronization phenomenon. In coupled van der Pol oscillators, synchronization occurs when a system of two or more oscillators interacts. Our investigation demonstrates that introducing the squeezing Hamiltonian in two coupled van der Pol oscillators enhances nonclassical effects, increases quantum correlations, and improves the robustness of synchronization dynamics. This was evidenced through the analysis of the Wigner function and power spectrum, showing significant improvements compared to systems without squeezing.
\end{abstract}

\maketitle
\section{Introduction}
\label{introduction}
Synchronization is the phenomenon where a set of oscillating systems adjust their rhythms or behaviours to become harmonized or coordinated with each other. In other words, it is the process by which different systems or components become aligned in time or frequency. For example, in the context of coupled oscillators \cite{1,2}, synchronization occurs when the phases or frequencies of the oscillators become correlated. This means that the individual oscillators adjust their dynamics to achieve a common phase or frequency relationship, leading to a coherent and collective behaviour.  When multiple oscillators or waves are synchronized, they also exhibit a coordinated behaviour, such as in the case of synchronized pendulum clocks or fireflies flashing in unison. Synchronization can occur in various domains, not only in physics but also in biology, and engineering \cite{3,4,5,6}.

Studies have explored phase-locking in driven quantum self-sustained oscillators \cite{7,8,9,10,11,12,13} as well, and it has been observed that multiple interacting oscillators can adapt their phase alignment in a way reminiscent of classical systems \cite{14,15,16,17,18,19,20,21} . Strong nonlinear damping rates \cite{22} are essential in these proposed configurations for attaining the quantum realm. This helps in attaining steady states with low average particle numbers. However, in experiments conducted with  ``micromechanical" \cite{23,24} , ``nanomechanical resonators" \cite{25,26,27}, and ``optomechanical oscillators" \cite{28}, a common challenge emerged: the systems under study were highly excited. As a result, our capacity to witness genuine quantum phenomena was constrained.

Recent advancements have introduced novel approaches for self-sustained oscillators to operate deeply within the quantum realm \cite{29,30}. In this innovative context, the influence of quantum fluctuations becomes more evident, challenging the system's synchronization to an external drive through the introduction of a distinct form of phase diffusion \cite{11}. Upon an initial overview, systems operating close to their ground state may seem inappropriate for studying synchronization. Nevertheless, we demonstrate that this is not necessarily the case. The challenges arising from increased noise due to quantum fluctuations can be mitigated by utilizing an additional fundamental quantum phenomenon, namely ``squeezing" \cite{31,32}.

 Through the substitution of the external drive with a squeezing Hamiltonian, we successfully counteract the detrimental influence of noise, thereby unlocking the prospect of witnessing quantum synchronization within the depths of the quantum domain in both single forced van der Pol (vdP) oscillator \cite{33} and in coupled vdP oscillators. When we add a squeezing Hamiltonian along with an external drive, synchronization significantly improves, and this improvement is more noticeable with the squeezing Hamiltonian than with the external drive alone. This holds true when we examine a harmonic drive and introduce squeezing separately. This deliberate distinction emphasizes the unique roles that each of these components plays in the quantum realm, clearly establishing squeezing as the primary factor in aligning the phase and frequency in both single forced vdP oscillator and in coupled vdP oscillators.
 This paper explores the transformative impact of squeezing, highlighting its ability to produce (i) improved synchronization, (ii) steady-state power spectrum across frequencies when the system is stable and the driving and system frequencies are closely matched, and (iii) the steady states exhibit authentic nonclassical traits like enhanced quantum correlations, nonclassical coherence, and reduced decoherence effects. Squeezing can strengthen correlations between the coupled VdP oscillators, leading to more synchronized behavior, while also sustaining nonclassical coherence by preserving superposition states. Furthermore, it helps to reduce decoherence, allowing these nonclassical features to persist for longer durations.

Our aim is to enhance synchronization in coupled oscillators by introducing squeezing parameters to mitigate quantum noise and exhibit non-classical behaviour. The introduction of squeezing parameters is expected to address quantum noise issues and manifest non-classical behaviour in the coupled oscillator system.

We organize our demonstration as follows. In Sec 2, we focus on examining how squeezing influences two coupled vdP oscillators, exploring its implications on frequency entrainment. In Sec 3, we give an experimental realization using trapped ions and a optomechanical system. In Sec 4, we provide an overview about the results we obtained in the previous sections.
\section{Two coupled van der Pol oscillators with squeezing}
When we have two vdP oscillators that are coupled, it means that the motion of each oscillator depends not only on its own state but also on the state of the other oscillator. Reactive coupling in vdP oscillators would involve a type of interaction where the exchange of energy is characterized by oscillatory or wave-like behaviour. This might correspond to a coupling mechanism that allows the oscillators to exchange energy without leading to a net gain or loss. The energy transfer is associated with the reactive components of the vdP oscillators. Dissipative coupling in vdP oscillators implies an interaction that results in a net dissipation or loss of energy from the coupled oscillators. This could occur if there are elements in the coupling mechanism that introduce damping or resistive effects, leading to a gradual reduction in the overall energy of the coupled system. In this section, we discuss the effects of squeezing and coupling parameters in these two types of coupled vdP oscillators.
\subsection{\textbf{Reactively coupled vdP oscillators}}The master equation for reactively coupled vdP oscillators can be expressed as follows:
\begin{eqnarray}
     \dot{\rho}&=&-i[\Delta_1 \hat{a}_1^{\dagger} \hat{a}_1 +\Delta_2 \hat{a}_2^{\dagger} \hat{a}_2+V[\hat{a}^\dag_1 \hat{a}_2 + \hat{a}_1\hat{a}^\dag_2],\rho] \nonumber \\
     &&+\gamma^{(1)}_{1} \mathcal{D}\left[\hat{a}_1^{\dagger}\right] \rho
     +\gamma^{(2)}_{1} \mathcal{D}\left[\hat{a}_2^{\dagger}\right] \rho
     +\gamma^{(1)}_{2} \mathcal{D}\left[\hat{a}_1^{2}\right] \rho  \nonumber \\
     &&+\gamma^{(2)}_{2} \mathcal{D}\left[\hat{a}_2^{2}\right] \rho .
\end{eqnarray}
Here, the term $D[\hat{O}]\rho = \hat{O}\rho \hat{O}^\dag - [{\hat{O}^\dag \hat{O},\rho}]/{2}$ represents Lindblad evolution. Let $\hat{a}_1(\hat{a}_2)$ and $\hat{a}_1^\dag(\hat{a}_2^\dag)$ be the annihilation and the creation operators of the two oscillators. The parameters $\gamma^{(1)}_{1}(\gamma^{(2)}_{1})$ and $\gamma^{(1)}_{2}(\gamma^{(2)}_{2})$ denote the rates of linear excitation and nonlinear dissipation and $V$ is the coupling strength of the two vdP oscillators, respectively.

The application of squeezing to the two coupled vdP oscillators \cite{34,35,36} involves the incorporation of a squeezing Hamiltonian for a degenerate parametric down conversion process. This Hamiltonian is represented as $ H_{sq}=i \sum_{j=1}^2 \chi^{(2)}\left(\hat{a}_j^{2} \hat{c}_j^{\dagger}-\hat{a}_j^{\dagger 2} \hat{c}_j\right)$, where $\hat{a_j}$ corresponds to the signal mode, $\hat{c}_j$ denotes the pump mode, and $\chi^{(2)}$ signifies nonlinear susceptibility of order two. The total Hamiltonian involving coupling and squeezing can be expressed as follows:
\begin{eqnarray}\label{Htot}
   \hat{H}_{tot} &=&\Delta_1 \hat{a}_1^{\dagger} \hat{a}_1+\Delta_2 \hat{a}_2^{\dagger} \hat{a}_2+V[\hat{a}^\dag_1 \hat{a}_2 + \hat{a}_1\hat{a}^\dag_2] \nonumber \\
   &&+i\sum_{i=1}^2 \eta_i\left(\hat{a}_i^{2} e^{-i \theta}-\hat{a}_i^{\dagger 2} e^{i \theta}\right).
\end{eqnarray}
Expressing the squeezing parameter as $\eta=\chi^{(2)} \lambda$, a quantity that encompasses both the second-order nonlinear susceptibility and the strength of the pump mode, the master equation takes on a comprehensive form. This inclusivity accounts for the standard constituents associated with linear excitation and nonlinear dissipation. The all-encompassing master equation has the form
\begin{eqnarray}\label{rho}
    \dot{\rho}&=&-i\left[\hat{H}_{\mathrm{tot}}, \rho\right]+ \gamma^{(1)}_{1} \mathcal{D}\left[\hat{a}_1^{\dagger}\right] \rho+\gamma^{(2)}_{1} \mathcal{D}\left[\hat{a}_2^{\dagger}\right] \rho \nonumber \\
     &&+\gamma^{(1)}_{2} \mathcal{D}\left[\hat{a}_1^{2}\right] \rho +\gamma^{(2)}_{2} \mathcal{D}\left[\hat{a}_2^{2}\right] \rho .
\end{eqnarray}
Equation (\ref{rho}) exhibits two distinct regimes with contrasting characteristics. When $\eta=0$ and $V\neq0$, Eq. (\ref{rho}) simplifies to the well-known ordinary coupled vdP equation that has been commonly explored in the literature \cite{34}. On the other hand, when $\eta\neq0$ and $V=0$, an uncharted territory is unveiled, which we term $"squeezing-coupled~-~vdP"$  regime.
\subsection{\textbf{ Classical picture}}
To facilitate a clearer understanding of the equilibrium points in the dynamics regulated by Eq. (\ref{rho}), we initiate our analysis by deducing the classical equations of motion. Under the circumstance where the oscillators exhibits significant excitation $(\gamma_{1} \gg \gamma_{2})$, we can substitute the operators $\hat{a}_1(\hat{a}_2)$ with its corresponding averages denoted as $\langle\hat{a}_1\rangle = R_1 e^{i \phi_1}$ and $\langle\hat{a}_2\rangle = R_2 e^{i \phi_2}$. This substitution allows us to formulate the following set of coupled equations, that is
\begin{subequations}
  \begin{eqnarray}
    \dot{R_1}&=&\frac{\gamma^{(1)}_1}{2} R_1-\gamma^{(1)}_{2} R_1^{3}-V R_2 \sin (\phi_2 - \phi_1)  \nonumber \\
    &&-2 \eta R_1 \cos (2 \phi_1-\theta),\label{R1} \\
     \dot{R_2}&=&\frac{\gamma^{(2)}_{1}}{2} R_2-\gamma^{(2)}_{2} R_2^{3}-V R_1 \sin (\phi_2 - \phi_1)\nonumber\\
     &&-2 \eta R_2 \cos (2 \phi_2-\theta),\label{R2} \\
    \dot{\phi_1}&=&-\Delta_1+V\frac{R_2}{R_1} \cos (\phi_2 -\phi_1)\nonumber \\
    &&+2 \eta \sin (2 \phi_1-\theta), \label{phi1} \\
     \dot{\phi_2}&=&-\Delta_2 +V\frac{R_1}{R_2}\cos (\phi_2 -\phi_1) \nonumber \\
     &&+2 \eta \sin (2 \phi_2-\theta).\label{phi2} 
  \end{eqnarray}
\end{subequations}
In the above equations \eqref{R1} -\eqref{phi2}, $R_1$ and $R_2$ represent the amplitudes of two oscillators, while $\phi_1$ and $\phi_2$ represent their respective phases. 
It will be challenging to comprehend the behaviour of the two-dimensional nullcline plot as we need to depict the four terms $R_1$, $R_2$, $\phi_1$, and $\phi_2$.  
\begin{figure}[htbp]
  \centering
  \includegraphics[width=0.75\textwidth,,height=2.0in]{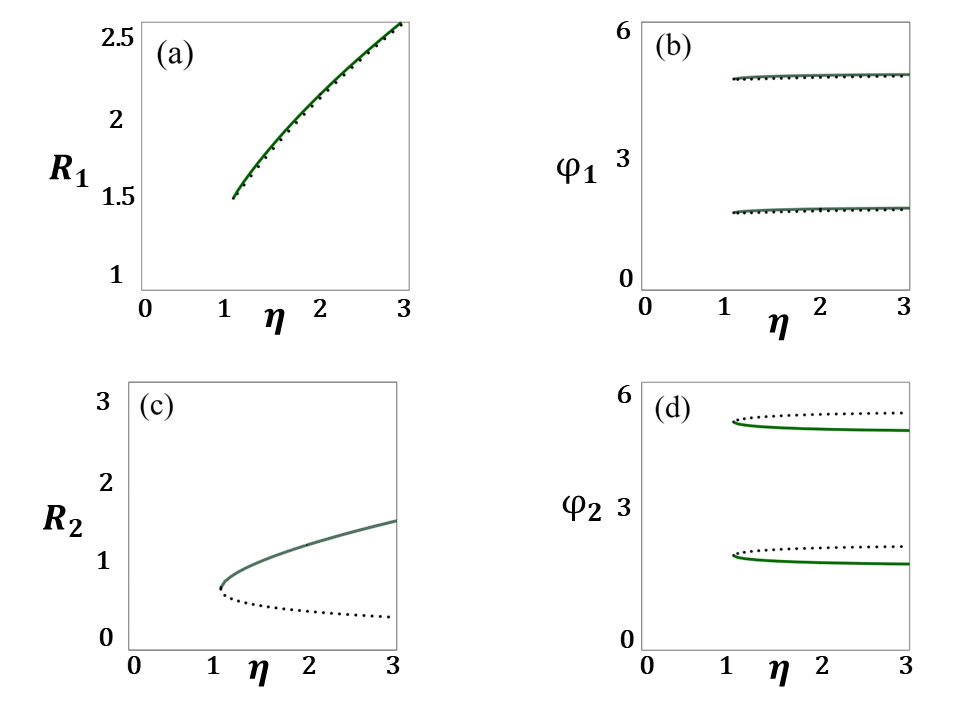}
     \caption{Bifurcation diagram of reactively coupled vdP oscillators for the values $\eta/\gamma^{(1)}_1$ = [1,3]. In these diagrams, figures (a) and (c) show the effect of $ \eta$ on the variables $R_1$, $R_2$ and figures (b) and (d) show the effect of $ \eta$ on the variables $\phi_1$ and $\phi_2$ respectively. The values of other parameters are $ V/ \gamma^{(1)}_1 = 1$, $\Delta_1/\gamma^{(1)}_1 = 1$, $\Delta_2/\gamma^{(1)}_1 = 1$, $\theta = \pi/4$, $\gamma^{(1)}_2/\gamma^{(1)}_1$ = 3 and other $\gamma$ values are 1.}
     \label{fig:1}
 \end{figure}
To comprehend the classical behaviour of these coupled oscillators, we have obtained the bifurcation diagrams using XPPAUT (AUTO) software \cite{37}. The bifurcation diagram with
respect to the squeezing parameter$ (\eta) $ is presented in all the variables $R_1$, $R_2$, $\phi_1$ and $\phi_2$
in Fig. \ref{fig:1}. The continuous green curves (colour line) in the figure represent stable steady state and the dashed black curves represent the unstable steady states. In this bifurcation diagram Fig. 1(a) - 1(d), we can observe the existence of two stable steady states with same $ { R^*_1}$ and ${R^*_2}$ values and two different ${\phi^*_1}$ and ${\phi^*_2}$ values and these states collide with unstable states represented by dashed black curves at $\eta$ nearly equal to 1 and results in saddle fixed point. Thus we can observe a saddle node type bifurcation in the system. Thus, in the classical case, the increase in the squeezing parameter stabilizes two different stable states with different phases $\phi_1$ and $\phi_2$ through saddle node bifurcation. More detailed bifurcation diagrams illustrating the impact of coupling strength across different detuning and coupling values are presented in the Appendix.

 In the quantum realm, the presence of squeezing introduces quantum coherence and correlations between the oscillators.  This quantum behaviour builds upon the classical understanding obtained through the analysis of bifurcation diagrams, providing a more subtle perspective that incorporates quantum effects. 
\subsection{\textbf{Quantum picture}}
\begin{figure}[htbp]
  \centering
    \includegraphics[width=\linewidth, height=7.5cm]{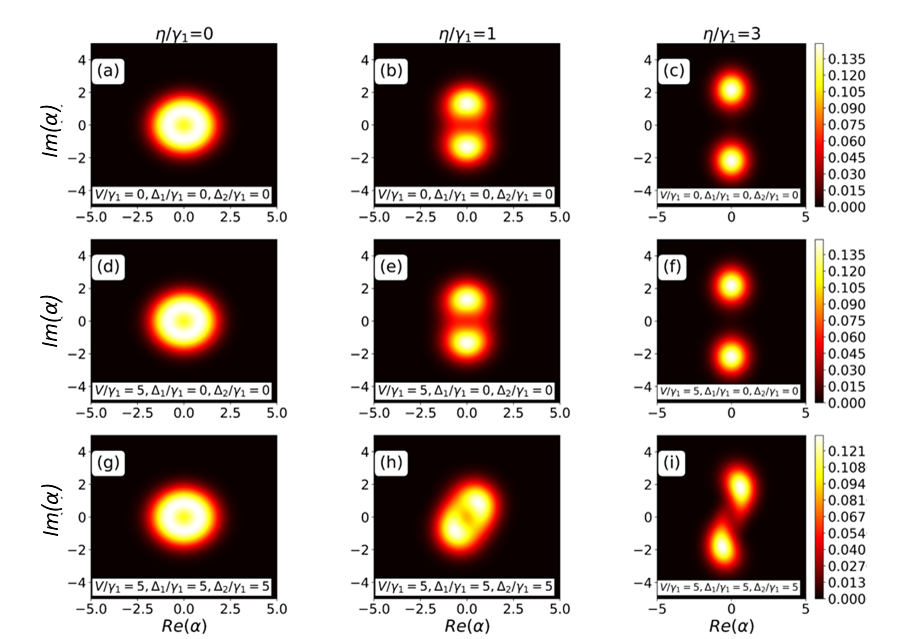}
  \caption{In the study of the Wigner function of reactively coupled vdP oscillator, we investigate three scenarios with varying $\eta/\gamma^{(i)}_1$ values, i=1,2. In the uncoupled vdP oscillators (a-c), squeezing creates a symmetric split when $V/\gamma^{(i)}_1 = 0$ and $\Delta_1/\gamma^{(i)}_1=\Delta_2/\gamma^{(i)}_1 = 0$. Moving to the coupled vdP oscillator (d-f), as $\eta$ increases for $V/\gamma^{(i)}_1 = 5$ and $\Delta_1/\gamma^{(i)}_1=\Delta_2/\gamma^{(i)}_1 = 0$, we observe symmetric bifurcation as in the case with $V/\gamma^{(i)}_1 = 0$. Lastly, in coupled oscillators with detuning (g-i), characterized by $V/ \gamma^{(i)}_1 = 5$ and $\Delta_1/\gamma^{(i)}_1=\Delta_2/\gamma^{(i)}_1 = 5$, detuning slightly disrupts symmetry but does not cause the lobe to vanish as in the case of the forced vdP oscillator. All these scenarios are set in a low excitation regime $\gamma^{(i)}_2/\gamma^{(i)}_1 = 3$, with squeezing applied along the position quadrature $\theta = 0$. }
\label{fig:2}
\end{figure}

The bifurcation phenomenon which arises from the classical solutions which is shown in Fig. \ref{fig:1} is notably recognizable in the form of steady state Wigner functions through numerical computation using QUTIP software \cite{38,39} which is shown in Fig. \ref{fig:2}.

From Fig. \ref{fig:2}, we can see that in the absence of squeezing, (columns (a)-(g)), the system retains its rotational symmetry, signalling the lack of coherence accumulation among the Fock states of individual oscillators and confirming the persistent diagonal nature of the reduced density matrix. In the presence of finite squeezing, see column (b)-(h), the Wigner function divides into two symmetrical lobes which is similar to the case of squeezing driven vdP oscillator. 

For a single oscillator without coupling, the quantum state is more susceptible to localization, leading to the disappearance of one lobe under large squeezing and detuning. However, for the coupled oscillator in the presence of finite detuning $\Delta$, this symmetry is not disrupted as in the case of single oscillator, when we further increase the squeezing, as seen in column (c)-(i), symmetry is well defined indicating the presence of strong coupling between oscillators. Coupling acts as a stabilizing factor that sustains the non-classical features of the quantum state in two coupled oscillators.
 
We now compare the frequency entrainment characteristics of a coupled vdP system with those of a system driven by squeezing. This comparison takes place within the realm of deep quantum behaviour. To investigate how the oscillator's frequency synchronizes with the coupling, we employ the observed frequency $\omega_{\text{obs}}$. This observed frequency is defined as the point where the power spectrum $S(\omega)$ reaches its highest value. When the oscillator's entrainment is subtle, the observed frequency $\omega_{\text{obs}}$ stays near the initial detuning $\Delta$. Conversely, when strong entrainment occurs, $\omega_{\text{obs}}$ progressively moves toward $\omega=0$. This shift signifies that the oscillations between the two oscillators are now closely aligned.

\begin{figure}[ht!]
	\centering
	\includegraphics[width=0.75\textwidth]{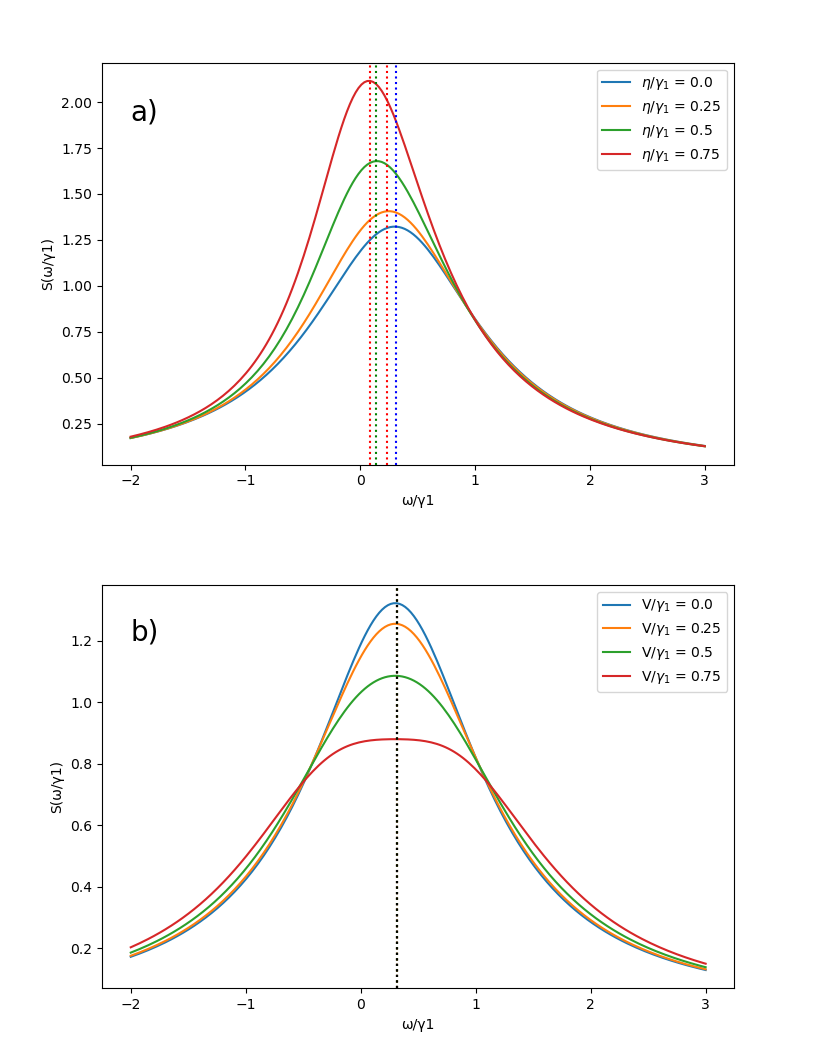}
	\caption{Investigating the entrainment of squeezing and reactively coupled vdP oscillators, Note that the ratio of dissipative processes is ${\gamma^{(i)}_2}/{\gamma^{(i)}_1} = 3$. In figures (a) and (b), we observe the power spectrum, denoted as $S(\omega/\gamma_1)$, when $\Delta_1 /{\gamma^{(i)}_1}={\Delta_2}/{\gamma^{(i)}_1}= 0.3$. In these plots, we can see that, upon increasing the parameter squeezing the system approaches zero detuning, while increasing the coupling has no effects on decreasing detuning.}
	\label{fig:3}
\end{figure}

When we examine the effects of a coupling, it becomes evident that there is a lack of significant frequency entrainment, even when the detuning $\Delta$ is exceptionally small. The underlying explanation lies in the fact that the influence exerted by the coupling is relatively weak, thereby lacking the strength required to overcome the inherent noise characteristics intrinsic to the vdP oscillator when compared to the squeezing parameter. The Fig. \ref{fig:3} provides a visual representation of this phenomenon, showcasing the behaviour of the power spectrum $S(\omega)$.

 Figure \ref{fig:3} (a) depicts plots for various squeezing parameter values, with the coupling set to zero. Notably, it illustrates that as the squeezing parameter increases, the detuning decreases, thereby leading to enhanced and robust synchronization. On the other hand, Fig. \ref{fig:3} (b) showcases plots for different coupling parameter values, with the squeezing parameter held at zero. In this case, the observation is that increasing the coupling parameter does not bring the detuning (close) to zero. This comparison suggests that squeezing has a more pronounced effect on synchronization when compared to the coupling parameter, as it leads to a stronger and more robust synchronization behaviour.

Upon comparing Wigner function figure given in Ref. \cite{31} and our Fig. (\ref{fig:2}), it is evident that in the squeezing-driven oscillator, an increase in squeezing results in the emergence of two equilibrium points. However, as the detuning is increased, one of these equilibrium points disappears, indicating the dominance of detuning in this scenario and the corresponding Arnold tongue behavior was already studied in \cite{31}. Conversely, in the case of the squeezing-coupled oscillator, an increase in squeezing leads to the presence of two equilibrium points. As detuning increases, these equilibrium points enter a metastable state, where they persist but become slightly deformed rather than one disappearing completely, suggesting that squeezing and coupling plays a dominant role in this context. This gradual deformation with increasing detuning is reminiscent of Arnold tongue behaviour, where synchronization regions shift and reshape rather than abruptly disappearing, highlighting the interplay between detuning and coupling strength in determining system stability.
\subsection{\textbf{ Dissipatively coupled vdP oscillators with squeezing}} 
The master equation for dissipatively coupled vdP oscillators \cite{40,41} can be written as,
\begin{eqnarray}
     \dot{\rho}&=&-i\sum_{i=1}^2\left [\Delta_i \hat{a}_i^{\dagger} \hat{a}_i,\rho \right]+ V\mathcal{D}\left[\hat{a}_1 - \hat{a}_2\right] \rho \nonumber \\
    && +\sum_{i=1}^2\gamma^{(i)}_{1} \mathcal{D}\left[\hat{a}_i^{\dagger}\right] \rho 
      +\sum_{i=1}^2\gamma^{(i)}_{2} \mathcal{D}\left[\hat{a}_i^{2}\right] \rho.
\end{eqnarray}
The total Hamiltonian involving coupling and squeezing can be expressed as follows:
\begin{equation}
   \hat{H}_{tot} = \sum_{i=1}^2 \Delta_i \hat{a}_i^{\dagger} \hat{a}_i +i\sum_{i=1}^2 \eta_i\left(\hat{a}_i^{2} e^{-i \theta}-\hat{a}_i^{\dagger 2} e^{i \theta}\right).
\end{equation}
The master equation with this total Hamiltonian has the form
\begin{eqnarray}\label{rho2}
    \dot{\rho}&=&-i\left[\hat{H}_{\mathrm{tot}}, \rho\right]+ + V\mathcal{D}\left[\hat{a}_1 - \hat{a}_2\right] \rho \nonumber \\
    && +\sum_{i=1}^2\gamma^{(i)}_{1} \mathcal{D}\left[\hat{a}_i^{\dagger}\right] \rho 
      +\sum_{i=1}^2\gamma^{(i)}_{2} \mathcal{D}\left[\hat{a}_i^{2}\right] \rho.
\end{eqnarray}
Upon exploring both classical and quantum aspects of dissipatively coupled vdP oscillators with squeezing, we observed analogous behaviours to those found in reactively coupled oscillators with squeezing (results are shown in Appendix). The classical phase plane diagrams and Wigner function representations exhibit the same patterns observed in the reactively coupled scenario. This suggests that irrespective of whether the coupling between oscillators is reactive or dissipative, the influence of coupling and squeezing on the system's dynamics remains same.
\section{ Experimental Feasibility}
In this section we present two experimental setups one using trapped ions and the other using optomechanical system. To explain this setups we follow the methodology outlined in \cite{29}. Our approach based on ion traps, utilizes the oscillator mode ${\hat{a}}$ to represent the linearly damped motional degree of freedom of a trapped ion. Laser cooling methods, as discussed in \cite{42}, offer a viable approach for incorporating linear damping. This involves driving the internal degree of freedom of an ion using a standing-wave laser field with the rate of oscillation between two quantum states (Rabi frequency) that aligns with the resonance of the first blue sideband transition. In the regime characterized by the Lamb-Dicke condition, where the amplitude of the quantum oscillator is significantly smaller than the wavelength of the associated radiation, and within tightly confined trapping potentials, this arrangement establishes an undriven vdP oscillator. This characteristic is substantiated by the distinct steady-state Wigner function exhibiting ring-shape which is observed \cite{29}.

Squeezing can be introduced into the system through various means. One such approach involves the combined application of standing- and travelling-laser fields, as discussed in \cite{43}. Another method includes inducing an adiabatic reduction in the trap's spring constant, as shown in \cite{44}. Moreover, achieving squeezing involves exposing the ion to a pair of Raman beams with a frequency difference of $2\omega_d$, as detailed in \cite{45}.

Furthermore, the realization of a vdP oscillator is feasible within a system demonstrating second-order nonlinearity, as explained in \cite{46}. In this particular context, the system includes a high-quality factor membrane characterized by minimal mechanical dissipation. The application of a blue-detuned laser with a solitary mechanical frequency corresponds to the linear pumping Lindbladian. Simultaneously, the introduction of nonlinear damping is facilitated by utilizing a red-detuned laser with two mechanical frequencies. Generating an electric field gradient in close proximity to the membrane allows for the creation of the force driving the dynamics.

Electrical modulation of the spring constant at double the mechanical frequency, as indicated in \cite{47}, enables the generation of squeezing. This framework provides a versatile and controllable platform for exploring the dynamics of the vdP oscillator in different physical systems.

\section{Conclusion}
This paper presents the extension of the impact of a squeezing Hamiltonian in a single-driven oscillator  to encompass a system involving two coupled van der Pol oscillators \cite{31,32}. Additionally, our exploration delves into assessing the resilience of synchronization induced by the squeezing Hamiltonian when applied to two coupled oscillators. Further, we have investigated whether the synchronization benefits achieved by the squeezing Hamiltonian are preserved or altered in this setup.
Our results indicate that the presence of a squeezing Hamiltonian continues to enhance synchronization in the two coupled vdP oscillators system. Additionally, the squeezing Hamiltonian is shown to help counteract the negative effects of noise, supporting and keeping synchronization in the connected system. Our study also looked at how coupling and squeezing each affect synchronization by studying their impact on the power spectrum of the system. We also found that squeezing produces a strong synchronization than coupling in coupled systems. Finally, we explained two experimental setups to produce the linear dampings, nonlinear dampings, and squeezing experimentally.
In quantum optics, stronger synchronization in vdP oscillators can be utilized for generating entangled photon pairs with improved temporal correlations, enhancing the efficiency of quantum information processing and quantum communication protocols \cite{48}. 
\section*{Appendix}
In this appendix, we present classical bifurcation diagrams corresponding to various values of detuning and coupling strength, observed in both dissipatively and reactively coupled oscillators. Additionally, we provide a detailed quantum perspective on dissipative coupling.
\section*{A. Classical bifurcation diagrams}
\begin{figure}[ht!]
  \centering
    \includegraphics[width=\linewidth, height=7.5cm]{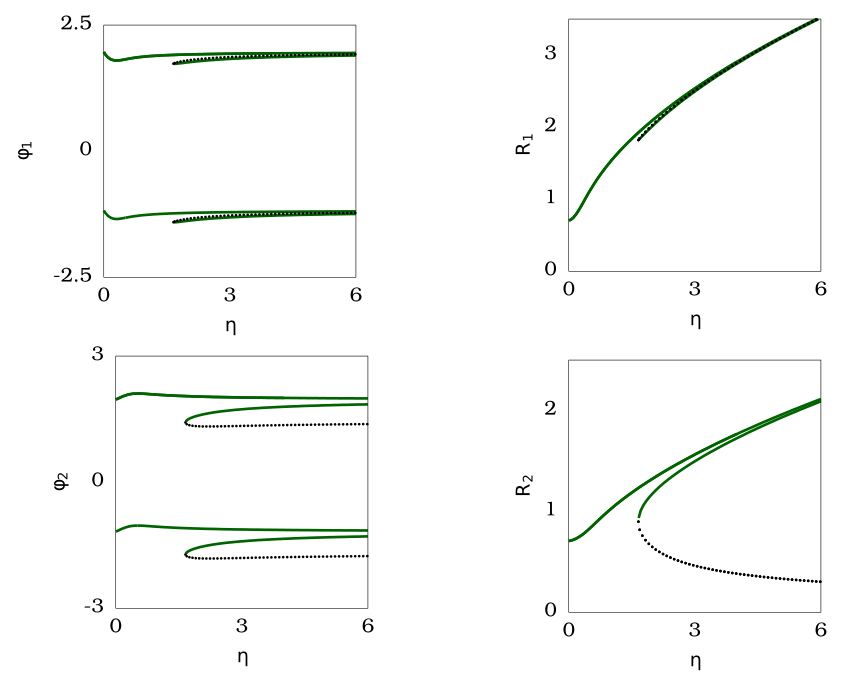}
  \caption{Bifurcation diagram for $\Delta_1$ = 1, $\Delta_2$ = 1, $\eta$ = [0,6] and V = 1}
\label{fig:4}
\end{figure}
\begin{figure}[ht!]
  \centering
    \includegraphics[width=\linewidth, height=7.5cm]{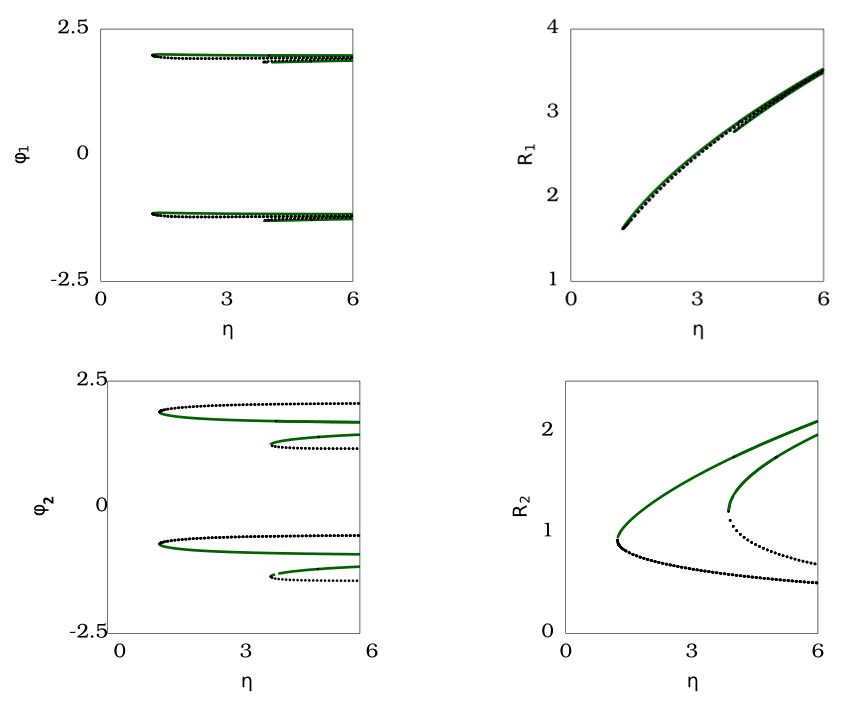}
  \caption{Bifurcation diagram for $\Delta_1$ = 1, $\Delta_2$ = 1, $\eta$ = [0,6] and V = 2 }
\label{fig:5}
\end{figure}
\begin{figure}[ht!]
  \centering
    \includegraphics[width=\linewidth, height=7.5cm]{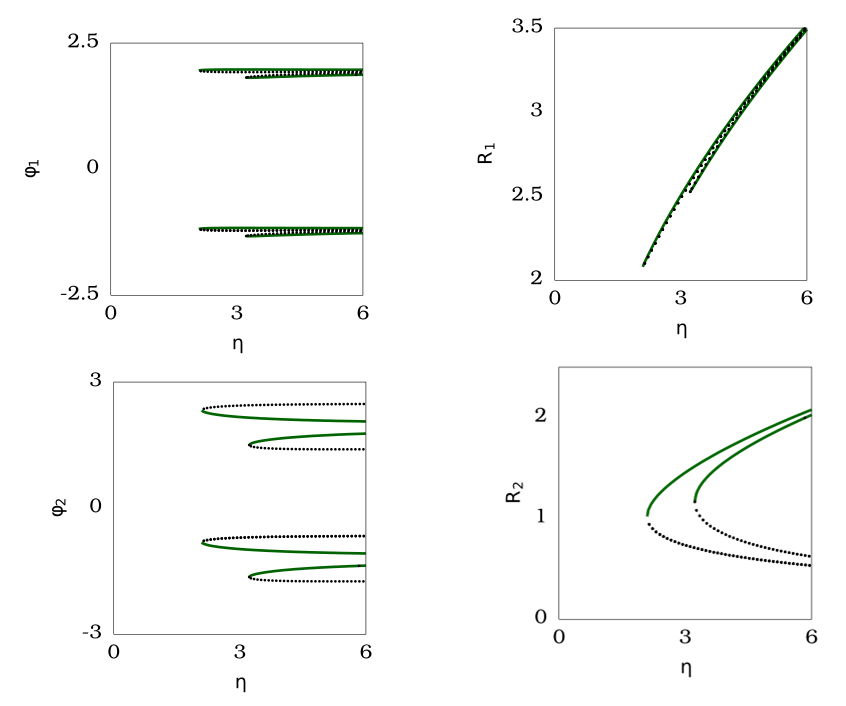}
  \caption{Bifurcation diagram with different detunings: $\Delta_1$ = 1, $\Delta_2$ = 2, V = 2, $\eta$ = [0,6] }
\label{fig:6}
\end{figure}
\subsection*{\textbf{A.1 Same detuning and different coupling strength ($\Delta_1=\Delta_2=1, V =[1,2]$ )}}
From Fig. \ref{fig:4}, it is evident that $\phi_1$ and $\phi_2$ exhibit two stable branches separated by an unstable region (dotted line), suggesting a transition between different phase-locked states and potential phase desynchronization. The amplitude $R_1$ increases smoothly with $\eta$, showing minimal instability. However, $R_2$ undergoes a clear bifurcation at a certain $\eta$, where one of its branches becomes unstable, indicating a loss of amplitude stability for the second oscillator at specific detuning values.

From Fig. \ref{fig:5}, a similar trend is observed for $V=2$ as in the case of $V=1$, with $\phi_2$ displaying even more complex multistability. While $R_1$ remains stable and grows more rapidly, suggesting that stronger coupling enhances the amplitude response, $R_2$ exhibits increased instability compared to $V=1$. More distinct branches emerge, indicating heightened multistability and the possible onset of amplitude death regions. This suggests that increasing the coupling strength does not enhance synchronization, as both phase and amplitude synchronization remain weak.

\subsection*{\textbf{A.2 Different detuning and same coupling strength
($\Delta_1=1, \Delta_2=2, V = 2$ )}}

\begin{figure}[ht!]
  \centering
    \includegraphics[width=\linewidth, height=7.5cm]{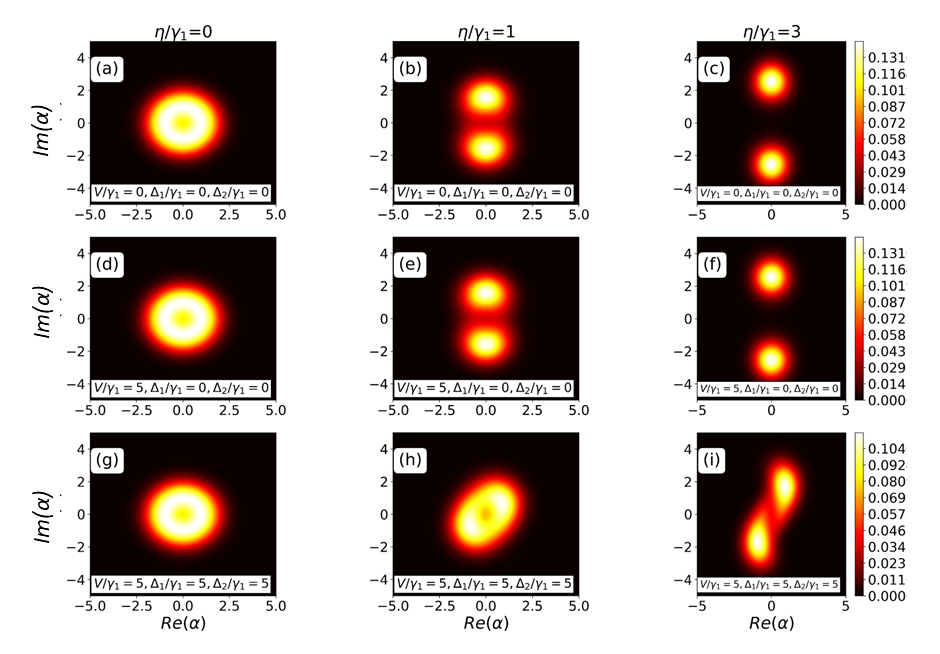}
  \caption{Illustrates the Wigner function of dissipatively coupled vdP oscillator, we investigate three scenarios with varying $\eta/\gamma^{(i)}_1$ values, i=1,2. In the uncoupled vdP oscillators (a-c), squeezing creates a symmetric split when $V/\gamma^{(i)}_1 = 0$ and $\Delta_1/\gamma^{(i)}_1=\Delta_2/\gamma^{(i)}_1 = 0$. Moving to the coupled vdP oscillator (d-f), as $\eta$ increases for $V/\gamma^{(i)}_1 = 5$ and $\Delta_1/\gamma^{(i)}_1=\Delta_2/\gamma^{(i)}_1 = 0$, we observe symmetric bifurcation as in the case with $V/\gamma^{(i)}_1 = 0$. Lastly, in coupled oscillators with detuning (g-i), characterized by $V/ \gamma^{(i)}_1 = 5$ and $\Delta_1/\gamma^{(i)}_1=\Delta_2/\gamma^{(i)}_1 = 5$, detuning slightly disrupts symmetry but does not cause the lobe to vanish as in the case of the forced vdP oscillator. All these scenarios are set in a low excitation regime $\gamma^{(i)}_2/\gamma^{(i)}_1 = 3$, with squeezing applied along the position quadrature $\theta = 0$. }
\label{fig:7}
\end{figure}
\begin{figure}[ht!]
  \centering
    \includegraphics[width=\linewidth, height=5.5cm]{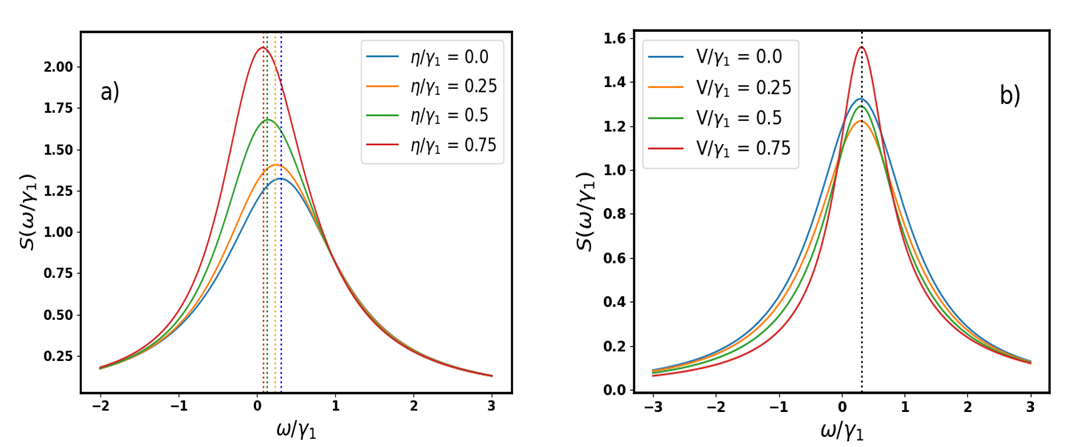}
  \caption{Power spectrum of dissipatively coupled oscillators for various squeezing (a) and coupling strengths (b). The effect of increasing squeezing reduces detuning, promoting synchronization, whereas increasing coupling strength does not significantly decrease detuning between oscillators.}
\label{fig:8}
\end{figure}
The bifurcation diagrams Fig. \ref{fig:6} reveal that different detuning values disrupt synchronization, leading to phase bistability and multistability $(\phi_1,\phi_2)$, where oscillators transition between distinct phase-locked states. While the first oscillator $R_1$ remains amplitude-stable, the second oscillator $R_2$ exhibits bifurcations and instability, indicating sensitivity to detuning and possible amplitude death. Despite equal coupling strength, the system does not achieve full synchronization, as detuning enhances phase desynchronization and destabilizes amplitude dynamics, particularly for the second oscillator.

\section*{B. Quantum picture of dissipatively coupled oscillator}

Dissipatively coupled oscillators exhibit dynamics similar to those of reactively coupled oscillators, as seen in the Wigner plots (Fig. \ref{fig:7}), where increasing the squeezing parameter leads to bifurcation patterns analogous to the reactive case. The power spectrum analysis (Fig. \ref{fig:8}) further supports this observation, showing that increased squeezing reduces detuning, thereby promoting synchronization between oscillators. However, increasing the coupling strength alone does not effectively reduce detuning, indicating that synchronization is primarily influenced by squeezing rather than coupling strength. This suggests that dissipation-induced coherence plays a crucial role in phase locking in the presence of squeezing, while coupling strength has a limited impact on frequency alignment.

\section{Declaration of competing interest}
The authors declare no conflict of interest.

\section{ACKNOWLEDGEMENT}
   The work of MS forms part of a research project sponsored by DST-SERB, Government of India, under the grant CRG/2021/00248.

\bibliographystyle{unsrtnat}

\end{document}